\documentclass [12pt]{article}
\usepackage{amsmath,amssymb,cite}
\usepackage[english]{babel}
\usepackage[dvips]{graphicx}

\usepackage{indentfirst}
\begin{document}
\begin{flushright}
KSU-09-2005\\
\end{flushright}

\begin{center}
{\large{\bf Modified GR and Helium Nucleosynthesis}}\\
\bigskip
A.S. Al-Rawaf.
\bigskip

{\it Physics Department, College of Science, King Saud University

P.O.Box 2455, Riyadh, Saudi Arabia}\\
\end{center}

\begin{abstract}
We show that a previously proposed cosmological model based on general
 relativity with non vanishing divergence for the energy-momentum tensor
 is consistent with the observed values for the nucleosynthesis of helium for
 some values of the arbitrary parameter $\alpha$  presented in  this model.  Further more values
 of $\alpha$  can be accommodated if we adopt the Randall-Sundrum single brane
 model.
\end{abstract}

Motivated by the desire to seek a solution to the entropy problem
within standard classical cosmology we have previously proposed a
simple modification of GR; namely relaxation of the condition of
the vanishing of the divergence of the energy-momentum tensor
\cite{1}
\begin{equation}\label{1}
  T^{\mu\nu};\mu=0
\end{equation}

This can be motivated by noting that the assumptions that lead to
(\ref{1}) are all questionable, including the principle of
equivalence \cite{2}. Relaxation of (\ref{1}) does not upset the
success of general relativity either in or outside cosmology. It
does, therefore, appear that the covariant conservation has not
been specifically tested by observation. The construction of a
theory in which (\ref{1}) does not necessarily hold would thus
provide an opportunity for testing this condition. In fact the
relaxation of this condition, introduces a single arbitrary
constant, $\alpha$ , where $ 0<\alpha  < 1$. The field equations
in this model become \cite{1}:
\begin{equation}\label{2}
  R_{\mu\nu}-\frac{1}{2}\gamma R g_{\mu\nu}=-K T_{\mu\nu}
\end{equation}

Where:
\begin{equation}\label{3}
\gamma=\frac{2-\alpha}{3-2\alpha} ~,~ K=\frac{8\pi G}{\alpha}.
\end{equation}

Imposing the Bianchi identity one obtains:
\begin{equation}\label{4}
  T^{\mu\nu};\mu =-\frac{1}{2}(1-\alpha) T,\nu
\end{equation}

Standard GR has  $\alpha = 1$.

 The vacuum field equations remain the same as in standard GR.
 Consequently the crucial tests of GR and the important analytic features
  (such as the existence of singularities and black holes) are maintained.
  The Field equations (\ref{2}) are also the same as the standard field equations
  for systems with traceless energy-momentum tensors, except that $G$ is replaced By $G/\alpha$.
    This shift in the effective value of $G$ to $G/\alpha$  in the early universe is expected
    to change the prediction of the primordial nuleosynthesis. In this paper we use a simple,
     approximate and semi analytical method to check the consistency of this model with the
      observed values for the nucleosynthesis of helium.  We show that at least for some values of
       $\alpha$ the model is consistent with the observed values. This problem was investigated in ref.
       \cite{3} using a different approach.

The modified filed equations, in the radiation dominated era in
this model become:
\begin{equation}\label{5}
 \bigg(\frac{\dot{R}}{R}\bigg)^{2}=H^{2}=\frac{8\pi G}{3 \alpha} \rho_{r}
\end{equation}

This represent an increase in density as related to standard
model.

The primordial production of $^{4}$He is controlled by a
competition between the weak interaction rates and the expansion
rate of the universe.  As long as the weak interaction rates are
faster than the expansion rate, the neutron-to-proton $\big(
\frac{n}{p}\big)$ ratio tracks its equilibrium value.  Eventually
as the universe  expands and cools, the expansion rate comes to
dominate and $ \frac{n}{p}$ essentially freezes out at the so
called freeze out temperature. However the nucleosynthesis chain
which begins with the formation of deuterium through the process $
p+n\rightarrow D+Y $  is delayed past the point where the
temperature has fallen below the deuterium binding energy $E_B$
since there are many photons in the exponential tail of the photon
energy distribution with energies  $E > E_B$ despite the fact that
the average photon energy are less than $E_B$  . (see for example
\cite{4}).

The increase in density is expected to modify the expansion of the
universe during nucleosynthesis by modifying both the freeze out
temperature and the time available for the decay of neutron. These
two factors will lead to an increase in the nucleosynthesis of
Helium.

Equation (\ref{5}) can be written as :

\begin{equation}\label{6}
 \bigg(\frac{\dot{R}}{R}\bigg)^2=\frac{8\pi G}{3}\rho_r
+\frac{8\pi G}{3} \frac{1-\alpha}{\alpha}\rho_r
\end{equation}

 The perturbation in density presented in Equation. (\ref{6}) can be
taken as a modification of the gravitational constant $\delta G$,
where
\begin{equation}\label{7}
  \delta G = \frac{1-\alpha}{\alpha} G
\end{equation}

Use of the conservation equation for radiation with equation
(\ref{6}) leads to:

\begin{equation*}
  t_m =\frac{t_s}{\alpha^\frac{1}{2}}
\end{equation*}

Where $t_m$  is the modified time from freeze out to start of
nucleosynthesis and  $t_s$ is the time calculated from Standard
Model without the extra energy.

 The calculation of primordial
element abundances is a highly nonlinear problem with many coupled
nuclear reactions, and requires a numerical analysis.  There are
many different numerical codes for doing this calculation starting
with Wagoner \cite{5}.  They mainly differ in the different
factors they include in their calculations and particularly the
different values they use for the neutron half life
\cite{6,7}.Here we only present a simplified and approximate
method. In this approximation, the primordial helium abundance
$Y_p$ is given by:
\begin{equation}\label{8}
  Y_p=\bigg(\frac{2N_n}{N_n+N_p}\bigg)_F \exp [ -\lambda
  (t_m-t_F)]
\end{equation}
\begin{equation}\label{9}
  \cong \bigg(\frac{2N_n}{N_n+N_p}\bigg)_F \exp (-\lambda t_m)
\end{equation}

Where $F$ represent freeze out and we ignored  $t_F$    because it
is of two order less than $t_m$
 Equation (\ref{9}) can be written as
\begin{equation}\label{10}
  Y= \bigg(\frac{2x}{1+x}\bigg)_F \exp( \frac{-\lambda t_s}{\sqrt{\alpha}})
  ~~~\text{where}~~~   x =\frac{N_n}{N_p}
\end{equation}

And thus
\begin{equation}\label{11}
  \delta Y = \frac{2 \delta x}{(1+x)^2} \exp(- \frac{\lambda t_s}{\sqrt{\alpha}})
\end{equation}

 $x =\frac{N_n}{N_p}=\exp\big(\frac{-1.5}{T_{f_{10}}}\big)$  , where  $T_{f_{10}}$ is the freeze out temperature
 $\bigg(T_n=\frac{T}{10^n K}\bigg)$
    and is determined by equating the Hubble constant $H$ to the weak reaction rate $\eta$  for  $n \leftrightarrow p$
     conversions. Now,
\begin{equation*}
 H \propto G^{1/2} T_{f_{10}}^2 ~~ \text{and}~~   \eta \propto
T_{f_{10}}^5
\end{equation*}

So that: $ T_{f_{10}}^3 = G^{1/2}$

%$$e^{-\frac{\lambda t_s}{\sqrt{\alpha }}} $$

\begin{equation}\label{12}
\therefore ~~~~ \delta Y = -\frac{(x\ln x) e^{-\frac{\lambda
t_s}{\sqrt{\alpha }}}} {3(1+x)^2} \frac{\delta G}{G}
\end{equation}

which from equation (\ref{7}) becomes:

\begin{equation}\label{13}
\delta Y = -\frac{(x\ln x)e^{-\frac{\lambda t_s}{\sqrt{\alpha }}}}
{3(1+x)^2} (\frac{1-\alpha}{\alpha})
\end{equation}

Putting $x=0.14$ (From $Y=0.246 $ \cite{6}), $\lambda = 78\times
10^{-5}$, $t_s\cong 120s $,  and if we take a value for $\alpha
=0.9$, equation (\ref{13}) gives:
\begin{equation*}
  \delta  Y=0.0071
\end{equation*}

which is quite within the margin of difference between measured
and observed values.  This value, as expected will increase, for
smaller values of $\alpha $  .

\paragraph{\bf Discussion}:

Big bang nucleosynthesis (BBN) is one of the most sensitive
available probes of physics beyond the standard model. The $^4He$
abundance in particular, has often been used as a sensitive probe
of new physics $[8 -12]$. This is due to the fact that nearly all
available neutrons at the time of BBN end up in $^4He$ and the
neutron-to-proton ratio is very sensitive to the competition
between the weak interaction rate and the expansion rate.
Observations of light element abundances have improved
dramatically over the past few years. The recent all-sky,
high-precision measurements of microwave background anisotropies
by WMAP has opened the possibility for new precision analysis of
BBN. Using BBN prediction gives a powerful constraint over the
various cosmological parameters. The possibility of the physical
'constants' taking different values at different times in the
universe history has recently received much attention with the
apparent observation that fine structure constant had a different
value in the distant past \cite{13}.  Variation of the
gravitational constant, that was originally started by Dirac
\cite{14} in his so called "Large Number" theory is now
constrained by the observations of the light element abundances
[15-18].

Copi et al \cite{16} uses the recent measurements of the
primordial deuterium abundances $(D/H)$ in
            conjunction with the WMAP determination of    $\eta$   to set a
Limit for $G/G_0$, where $G$ is the value for the gravitational
constant during nucleosynthesis and $G_0$ its value at present.
They found that $G/G_0$ is constrained in the range $1.21$ to
$0.85$ at the $68.3\% $
 confidence level and between $1.43$ and $0.71$ at the $95 \%$ confidence level Assuming a simple
power dependence $G \thicksim t^{-x}$, $x$ was constrained to the
range$ -0.004 <  x  <0.005$ at the $68.3\%$ confidence level, $
-0.009< x < 0.01$ at the $95\%$ confidence level. In our model
taking $\alpha = 0.9$ gives $G/G_0 = 1.11$,
 which is within the range obtained by Copi et al .Further Cyburt et al \cite{17} using
 $^4He$ abundance set a more restrictive limit for $ \Delta G/G_0$ of $13 \%$.

Finally we mention that if one adopted the recent ideas of brane
model, where the universe is supposed to be embedded in a higher
dimensional bulk, particularly the Randal Sunundrun type II model
\cite{19}, one could accommodate
 more smaller values for $\alpha$ .The predicted expansion rate in the early universe can be
 significantly modified, particularly with the presence of a dark radiation term \cite{20,21},
 which may take negative values and can contribute as high as $27\%$ of the background photon energy density \cite{22}.

\paragraph{Acknowledgements}
The author is grateful to the referee for raising some points that
led to the improvement Of the paper, and for drawing attention to
reference \cite{16}.

\bibliographystyle{unsrt}

\end{document}